\documentstyle[12pt,epsbox]{article}
\begin{document}

{\bf \Large Search for Hot Gas in the Local Group with ASCA}

\vspace{20pt}

Satoko OSONE$^{1,2}$ and Kazuo MAKISHIMA$^{2}$ and Keiichi MATSUZAKI$^{3}$ and

Yoshitaka ISHISAKI$^{4}$, Yasushi FUKAZAWA$^{5}$

\vspace{10pt}

[1] The Institute of Cosmic Ray Research, 5--1--5, Kashiwa-no-Ha, Kashiwa City, Chiba  277--8582

 [2]Department of Physics, University of Tokyo, 7--3--1 Hongo, Bunkyo-ku, Tokyo 113--0033

[3] The Institute of Space and Astronautical Science, 3--1--1, Yoshinodai, Sagamihara-shi, Kanagawa, Tokyo 229--8510

[4]  Department of Physics, Tokyo Metropolitan University, 1--1 Minami-Osawa, Hachiojishi, Tokyo 192--0397

[5] Department of Physics, Faculty of Science, Hiroshima University,1--3--1 Kagamiyama, Higashi-hiroshimashi, Hiroshima 739-8526

KeyWords[galaxies:Local Group --- cosmology: cosmic microwave background --- Xrays: diffuse background]

%\maketitle

\def\dfrac#1#2{{\displaystyle\frac{#1}{#2}}}

\begin{abstract}

 An X-ray study was made to examine whether some part of the soft X-ray background is coming from hot gas in the Local Group. 
For this purpose, four consecutive pointings were made with ASCA toward a sky region between M 31 and M 33, which is close to the direction of the center of the Local Group. By comparing the X-ray surface brightness in this sky direction with that in another blank sky region near the north equatorial pole, an upper limit on any soft excess X-ray background was determined to be 2.8$\times$10$^{-9}$ erg cm$^{-2}$ s$^{-1}$ sr$^{-1}$ with a 90\% confidence level statistical error. Assuming an optically thin thermal bremsstrahlung  energy spectrum (Raymond-Smith model) for a temperature of 1 keV and a $\beta$-model electron density distribution for a core radius of 100 kpc for the X-ray halo, the upper limit of the central plasma density was obtained to be 1.3$\times$10$^{-4}$ cm$^{-3}$. The plasma column density is too low to contribute significantly to the observed quadrupole anisotropy in the cosmic microwave background.
\end{abstract}

\section{Introduction}

\parindent=10pt It is well known that clusters of galaxies and elliptical galaxies contain an X-ray emitting hot gas. From X-ray observations with  ROSAT (Pildis et al. 1995; Mulchaey et al. 1996; Mulchaey, Zabludoff 1998) and ASCA (Fukazawa et al.\ 1996; Davis et al.\ 1999; Takahashi et al. 2000), it has become clear that groups of galaxies which are rich in elliptical galaxies also contain  hot gas having a temperature of $\sim$1 keV\@. Then, how about the Local Group? Because the Local Group is not rich in elliptical galaxies, we may not expect much hot gas in it. Also, X-ray emission from such a halo would be rather difficult to observe because such a hot gas in the Local Group must be widely spread and we would be inside it. Nevertheless, hot gas in the Local Group would have significant implications for X-ray Background (XRB) and Cosmic Microwave Background (CMB).

\parindent=10pt The origin of the XRB is mainly considered to be an accumulation of unresolved point sources (Fabian, Barcons 1992). In the 0.5--2 keV band, 70--80\% of the XRB has been resolved into point sources above 5.5$\times$10$^{-15}$ erg cm$^{-2}$ s$^{-1}$ with ROSAT (Hasinger et al. 1998) and Chandra (Mushotzky et al.\ 2000; Giacconi et al.\ 2001). However, XRB below 1 keV has an additional spatial complexity against an isotropic emission, as is clear in on all-sky map of ROSAT (Snowden et al.\ 1995). In addition, the XRB below 2 keV shows a spectral excess above the power-law component of photon index 1.4 that well describes the spectrum above 2 keV (Hashinger 1992; Gendreau et al.\ 1995; Ishisaki 1996; Chen et al.\ 1997; Miyaji et al.\ 1998; Parmar et al.\ 1999). The excess component of the XRB is contributed by optically thin thermal emission having a temperature of $\sim$0.1 keV from an ionized gas surrounding the solar system, and an emission of temperature of $\sim$0.7 keV associated with our Galaxy (e.g., Sidher et al. 1996; Parmar et al.\ 1999; Kuntz, Snowden 2000), as evidenced by line emissions from O{\tiny VII} and C{\tiny V} (Inoue et al.\ 1980; Rocchia et al.\ 1984) and a shadowing study of XRB with cold clouds (Kerp 1994; Snowden et al.\ 1997; Kerp et al.\ 1999). In addition to these components, there may also be thermal emission from the hypothetical Local Group halo contributing to the soft XRB.

\parindent=10pt A CMB quadrupole anisotropy of 6 $\mu$K has been detected with COBE (Bennett et al.\ 1994), and interpreted as a fluctuation made during inflation. Suto et al.\ (1996) have suggested that the quadrupole anisotropy might be partially contributed  by the Sunyaev--Zel'dovich (SZ) effect due to hot gas in the Local Group, though some authors claim that the effect should be negligible (Pildis, McGaugh 1996; Banday, Gorski 1996).

\parindent=10pt Using the ROSAT PSPC archives, Sidher et al.\ (1999) derived the electron core density of the Local Group to be (0.4$\pm$0.3)$\times$10$^{-3}$ cm$^{-3}$ at 0.17 keV, which is too low to cause any significant effect on the CXB or CMB.  However, Sidher et al.\ (1999) used XRB data far from the center of the Local Group toward the direction of M 31. Here, we report on our search for hot gas in the Local Group with ASCA (Tanaka et al.\ 1994) in the XRB spectra acquired in the direction toward the center of the Local Group, and present an upper limit on the electron density and the CMB anisotropy.

\section{Observation}
\parindent=10pt Assuming that the halo is densest toward the center of the Local Group, we searched for the Local Group X-ray halo by taking the difference in the XRB surface brightness between the direction of the center of the Local Group and the opposite direction. A calculation of galaxy kinematics (Peebles 1990) suggests that the center of the Local Group is situated in the middle between our galaxy and M 31. However, the pointing direction should be offset by several degrees from M 31 because M 31, itself, is a largely extended X-ray source. Therefore, we decided to observe around ($l,b$)=(123$^{\circ}$, $-28^{\circ}$), which is a middle point between M 31 and M 33, as shown in table 1 and figure 1. Here, we assume that the hot gas associated with the Local Group has a core radius larger than 40 kpc, which corresponds to this offset distance from the true center of the Local Group.  The galactic component of the XRB in this field can be neglected at around 1 keV\@. We performed four neighboring pointings with ASCA in order to reduce the risk of finding unexpected bright sources in the field of view. Hereafter, we call these observed fields the LGC field. The basic technique we used was subtraction of the GIS spectra acquired in the two directions, the LGC field and the LSS field.

\parindent=10pt For a reference sky region, we used the data of the ASCA Large Sky Survey (Ishisaki 1996; Ueda et al.\ 1998), covering a large sky area of 7.2 deg$^2$ near the north equatorial pole. These data are ideal as a reference, because its direction is about $113^\circ$ offset from the direction to the LGC. The X-ray surface brightness due to any hot gas in the Local Group towards this field was estimated to be 59\% of that in the LGC direction, assuming a $\beta$ model electron density distribution with a core radius of 500 kpc, as described in subsection 4.1. Hereafter, we call this reference sky the LSS field.

\section{ Data Analysis and Results}
\subsection{Data Screening and Non X-ray Background}
\parindent=10pt In comparing the surface brightness of the LGC and LSS, we used only data from the Gas Imaging Spectrometer (GIS; Ohashi et al.\ 1996) coupled to the X-ray telescope (XRT; Serlemitsos et al.\ 1995), primarily because of its larger solid angle than that of the Solid State Imaging Spectrometer (SIS; Burke et al.\ 1994). Moreover, the evaluation of the non X-ray Background (NXB) is well established and the stability of the detector performance has also been confirmed for the GIS, which is suitable for our purpose of searching very extended and faint X-ray emission. During observations of both the LGC and LSS, the GIS was operated in the PH normal mode.

\parindent=10pt We screened all of the GIS events taken in each observation, by first employing the standard event selection criteria; i.e,\ the f.o.v.\ elevation should be $\ge$ 5$^\circ$ (or 25$^\circ$) above a dark (or a sunlit) Earth rim, and the geomagnetic cutoff rigidity (COR) should be $\ge$ 6 GV\@. We performed a further event screening utilizing the GIS monitor counts H02 (i.e., H0$+$H2; see Ohashi et al.\ 1996; Makishima et al.\ 1996), Radiation-Belt-Monitor counts, and COR\@. This additional selection, called the H02 method, improved the NXB reproductivity, mainly by rejecting data affected by unexpected increases in the NXB counts, which are probably due to a concentration of charged particles on the satellite orbit. Further details of the screening procedure are described in Ishisaki (1996) and Ishisaki et al.\ (1997).

\parindent=10pt In order to determine the surface brightness of both fields, we basically collected events within a radius of 20' from the optical axes of GIS 2 and GIS 3, where the calibration of the GIS was accurate enough for our purpose. However, the obtained spectra also contains NXB events and X-ray emission from moderately bright point sources in the fields, which had to be subtracted (\S 3.2).

\parindent=10pt The NXB could be estimated by utilizing the GIS data acquired while the XTR was pointing to the night Earth, presuming that the night Earth emits no X-rays and that the atmospheric reflection of cosmic X-rays is negligible. For this purpose, a large amount of GIS NXB database was produced (Ishisaki 1996), by accumulating night-Earth data for 1400 ks. However, we did not conduct real-time NXB measurements during each on-source observation, and the NXB level is known to vary by as much as a factor of two or so depending, e.g., on the spacecraft location around the Earth. As a result, direct subtraction of the night Earth GIS spectrum from the on-source one introduces a large systematic error.

\parindent=10pt In order to overcome this problem, the so-called H02 method was developed (Ishisaki 1996). It utilizes the fact the one of the GIS monitor counts, called H02 (Ohashi et al.\ 1996), is tightly correlated with the instantaneous NXB normalization. We then sort the above mentioned NXB database into a sequence of NXB spectra according to the H02 counts, and combine them in reference to the actual H02 count distributions during the on-source exposure. The synthetic NXB spectrum created in this way can be subtracted from the on-source spectrum, after a correction for a long-term secular NXB increase by $\sim 5$\% yr$^{-1}$, which is probably due to an in-orbit build up of  long-decay radioactive isotopes (Ishisaki 1996).

\parindent=10pt Since the H02 method utilizes the long-exposure NXB database, the synthetic NXB spectra are relatively free from statistical errors. However, they are still subject to sporadic NXB variations, e.g., on day-by-day time scales, that are not clearly reflected in the H02 counts. This systematic effect limits the accuracy of the NXB subtraction (Ishisaki 1996; Makishima et al.\ 1996). The expected NXB reproducibility is limited to $\sim 7$\% in terms of the 90\% confidence systematic errors. In contrast, that in the LSS data is accurate to 3\%, because the sporadic NXB variations tend to average out over the 515 ks of on-source exposure for the LSS.

\subsection{Image and Removal of Point Sources}
\parindent=10pt In figure 2, we show an NXB-subtracted 0.7--7 keV image for the LGC, after smoothing with the detector-position-dependent point-spread function, as described by Takahashi et al.\ (1995). The corresponding LSS image is given by Ueda et al.\ (1998).

\parindent=10pt We can observe some point sources in the LGC image. There are also several moderately bright point sources in the LSS (Ishisaki 1996; Ueda et al.\ 1998). The brightest point source in the LGC has a 2--10 keV flux of 0.6$\times$10$^{-12}$ erg cm$^{-2}$ s$^{-1}$, while that in the LSS is 2$\times$10$^{-12}$ erg cm$^{-2}$ s$^{-1}$. These fluxes were derived by assuming a power-law spectra of photon index 1.7, together with the Galactic absorption.

\parindent=10pt In order to accurately compare the XRB brightness in the two fields, we must subtract point sources at the same flux threshold. With the source-masking method that Ishisaki (1996) developed, we removed point sources above a threshold flux of $2\times 10 ^{-13}$ erg s$^{-1}$ cm$^{-2}$ (2--10 keV). Each mask is circular, and its size is determined to cover up to a radius where the point spread function produced by the contained point source becomes 3\% of the XRB brightness. We also considered the vignetting effect. By this point-source exclusion, the effective area could be decreased by 23.5\% for the LGC and 24.7\% for the LSS.

\subsection{XRB and Local Group Halo}
\parindent=10pt After removing point sources, we summed up events from the two GIS detectors over the four fields of the LGC to obtain an XRB spectrum in the LGC direction. Exactly in the same way, we processed the LSS data. These spectra are presented in figure 3. In estimating the XRT effective area, we considered the effects of the point-source masks, as well as those of the vignetting and the stray light.

\parindent=10pt First, we fitted these spectra in the 2--10 keV band with a single power law, modified by the galactic absorption, 5.1$\times$10$^{20}$ cm$^{-2}$ for the LGC and 1.1$\times$10$^{20}$ cm$^{-2}$ for the LSS (given by EINLINE).  We obtained a photon index of 1.67 with a $\chi^2/{\rm d.o.f.}$ 1.17 for the LGC, and a photon index of 1.47 with a $\chi^2/{\rm d.o.f.}$ 1.20 for the LSS. When these best fits were extended into the entire 0.6--10 keV range, the fit a $\chi^2/{\rm d.o.f.}$ became 1.18 for the LGC and 3.82 for the LSS. A clear excess is seen in the energy range below $\sim$2 keV, particularly in the LSS spectrum, in agreement with the general XRB property (section 1). However, the soft excess appears to be less significant in the LGC fields. 

\parindent=10pt We next fitted these two GIS spectra with a two-component model, consisting of a hard power law and a soft power law, of the form $abs(E)*A E^{-\Gamma_{\rm h}}+B E^{-\Gamma_{\rm s}}$ in the 0.6--10 keV band. Here, $E$ is the X-ray energy, $A$ and $B$ are normalizations, $\Gamma_{\rm h}$ is the photon index of the hard power law, and $\Gamma_{\rm s}$ is the photon index of the soft power law. We applied the Galactic absorption $abs(E)=e^{-N_{\rm H}\sigma(E)}$, mentioned before, to the hard power-law component for each field. Here, $N_{\rm H}$ is a column density (cm$^{-2}$) and $\sigma(E)$ is a cross section of a photo absorption. When we left both $ \Gamma_{\rm h}$ and $ \Gamma_{\rm s}$ free, or fixed only $ \Gamma_{\rm h}$ at 1.4, the fit for the LGC had such large errors that we were unable to constrain the two components. Therefore, we fixed $ \Gamma_{\rm s}=$ 6.0 (after Ishisaki 1996) and left $\Gamma_{\rm h}$ free. The fitting results are presented in table 2, where the errors include both the statistical error (90\% confidence level) and systematic error of the H02 method (90\% confidence level) described in subsection 3.1. We show the fitted spectra and residuals in figure 3.

\parindent=10pt Thus, the soft XRB components in the LGC fields are in fact fainter than in the LSS filed. Allowing for errors, the 0.6--2 keV soft XRB flux in the LGC direction is at most 0.66 $\times10^{-8}$ erg cm$^{-2}$ s$^{-1}$ sr$^{-1}$, while that in the LSS direction is at least 0.38 $\times10^{-8}$ erg cm$^{-2}$ s$^{-1}$ sr$^{-1}$. By taking the differences between these two fluxes, we set a 90\% upper limit of 2.8$\times$10$^{-9}$ erg cm$^{-2}$ s$^{-1}$ sr$^{-1}$ to any excess in the soft XRB in the LGC direction above that in the LSS.

\parindent=10pt There is a systematic error for an absorption column, which may affect the soft power-law flux. We estimated this effect by assuming a systematic error of the $N_{\rm H}$ values to be $\pm$20\%. 
This affects the soft power-law flux of each XRB by $\pm$2\%, which has a negligible effect on the present upper limit. There could also be a contribution from the hot gas associated with our galaxy, and from the ionized gas surrounding the solar system. However, because the former is expected to be stronger on average in the LGC direction($b\simeq -30^{\circ}$) than in the LSS field($b\simeq 90^{\circ}$), a proper subtraction of this component would reduce the derived upper limit on the excess XRB flux in the LGC direction. We conservatively retained the upper limit on value mentioned above. The latter component, with a typical temperature of 0.1--0.15 keV, is too cool to contribute to the GIS band.

\section{Discussion}
\subsection{Limit to the Physical Parameters of Hot Gas}

\parindent=10pt We derive an upper limit on the electron density from an assumption that the spectrum of emission from the hot gas in the Local Group can be represented by optically thin thermal bremsstrahlung, specifically by a Raymond-Smith model(Raymond, Smith 1977), and its electron density distribution is represented by a $\beta$-model.

First, we translated the upper limit flux of the hot gas (table 2) into an observed normalization, $F_{\rm obs}$, of the Raymond-Smith model. The metal abundance was fixed to 0.3 solar based on results for other groups (Mulchaey et al.\ 1996; Fukazawa et al.\ 1996; Hickson 1997; Mulchaey, Zabludoff 1998). We fixed the plasma temperature in the range of 0.3--1.2 keV, stepping by 0.1 keV, to cover the typical values found in groups of galaxies (Pilidis et al.\ 1995; Fukazawa et al.\ 1996; Mulchaey et al.\ 1996; Mulchaey, Zabludoff 1998; Davis et al.\ 1999; Takahashi et al.\ 2000). Although the small velocity dispersion, about 60 km s$^{-1}$, of the Local Group might indicate a considerably lower temperature (e.g., $\sim$ 0.1 keV) for a hydrostatic hot gas, the GIS does not have sensitivity for such  cool emission. Therefore, our result is limited to any hot gas which is hotter than $\sim$ 0.3 keV.

\parindent=10pt Next, we must relate the observed normalization of the Raymond-Smith model, $F_{\rm obs}$, which is derived from the data, with physical and geometrical halo parameters.
 The normalization of a Raymond-Smith model ($F$) is generally given as
\vspace{10pt}

\begin{equation}
F= 10^{-14} \int \dfrac{ N_{\rm e}(r)^2 }{ 4\pi{\xi}^2 } dV.
\end{equation}
\vspace{10pt}

\parindent=0pt Here, 10$^{-14}$ is an arbitrary scale incorporated in the XSPEC model, $r$ (cm) is the distance from the center of the Local Group, $\xi$ (cm) is the distance to the X-ray emitting region, and ${\it N_{\rm e}}(r)$ (cm$^{-3}$) is the electron density at ${\it r}$. In reference to the geometry illustrated in figure 4, $r^2$ is given as

\vspace{10pt}
\begin{equation}
r^2 = \xi^2 - 2 x_{\rm 0} \xi {\rm cos}\theta + x_{\rm 0}^2,
\end{equation}
\vspace{10pt}

\parindent=0pt where $\theta$ is the angle between the center of the Local Group and the X-ray emitting region, and $x_{\rm 0}$ is the distance to the LGC. In a $\beta$ model, $N_{\rm e}(r)$ is given by

\vspace{10pt}
\begin{equation}
{\it N_{\rm e}}(r)=  {\it N}_{\rm 0} \left[ 1 + \left( \dfrac{r}{r_{\rm c}} \right)^2 \right]^{-2\beta/3}.
\end{equation}
\vspace{10pt}

\parindent=0pt Here, ${\it r_{\rm c}}$ (cm) is the core radius of the hot gas and ${\it N}_{\rm 0}$ (cm$^{-3}$) is the central electron density. 
We assume that the electron density obeys a $\beta$ model of $\beta=2/3$ based on the results for other groups (Pildis et al.\ 1995; Mulchaey et al.\ 1996; Mulchaey, Zabludoff 1998; Takahashi et al.\ 2000).
Then, the observed normalization, $F_{\rm obs}$ (~sr$^{-1}$~), for $\beta=$2/3, which is the difference of gas emission between the directions of the LGC and LSS, is written as 

\vspace{10pt}
\begin{equation}
F_{\rm obs} = \dfrac{10^{-14}}{4\pi} \times N_{\rm 0}^2 r_{\rm c} \dfrac{ \left( \dfrac{\pi}{2}+1 \right) }{2} \times
 \left[ 1 - \left( 1 + \left( \dfrac{ {\rm sin}\theta }{ \theta_{\rm c} } \right)^2 \right)^{-3/2} \right],
\end{equation}
%\begin{eqnarray}
% {\it F}_{\rm obs} &=& \dfrac{10^{-14}}{4\pi} \times N_{\rm 0}^2 r_{\rm c} \dfrac{ \left( \dfrac{\pi}{2}+1 \right) }{2} \times \nonumber \\
% & &  \left{ 1 - \left[ 1 + \left( \dfrac{ {\rm sin}\theta }{ \theta_{\rm c} } \right)^2 \right]^{-3/2} \right}.
%\end{eqnarray}
\vspace{10pt}

\parindent=0pt with $\theta_{\rm c}$ $\equiv$ ${\it r}_{\rm c}/{\it x_{\rm 0}}$, and $\theta=$ 1.97 rad being the angle between the LGC and the LSS.

\parindent=10pt By using equation (4), we translated the observed normalization of halo, $F_{\rm obs}$, into the quantity ${N_{\rm 0}}^2 r_{\rm c}$ for each plasma temperature.
We fixed $x_{\rm 0}=$ 350 kpc, which is  half the distance of M 31.
We took a core radius in the range of 100--500 kpc because other groups have a core radius of 40--400 kpc (Pildis, Mcgaugh 1996; Mulchaey , Zabludoff 1998; Mulchaey et al. 1996), and because we could not constrain any halo of core radius $\ll$ 40 kpc due to our observing strategy (section 2).
We present the result in figure 5. Thus, the upper limit on the central electron density of the Local Group is typically 1.3$\times$10$^{-4}$ cm$^{-3}$  for a core radius of 100 kpc and an assumed temperature of 1 keV\@. This value is considerably lower than those of X-ray emitting groups of galaxies (Mulchaey, Zabludoff 1998; Pildis et al.\ 1995). This is presumably because the Local Group is a poor group which does not contain any luminous elliptical galaxies. This result is consistent with the null detection (0.4$\pm$0.3) $\times$10$^{-3}$ cm$^{-3}$ with ROSAT of the 0.17 keV thermal emission (Sidher et al.\ 1999).

\subsection{Limit to the CMB Quadrupole Anisotropy}
\parindent=10pt The CMB quadrupole anisotropy, ${\it Q}_{\rm sz}$, is related to the halo parameters as 
\vspace{10pt}
%\begin{equation}
%{\it Q}_{\rm sz} = \dfrac{\sqrt5\pi\sigma_{\rm T} kT}{4fm_{\rm e}c^2} \times {\it N}_{\rm 0}{\it r}_{\rm c} \times 
 %\dfrac{ \left[ {\rm tan}^{-1}\left( \dfrac{ {\it x_{\rm 0}} }{  {{\it r}_{\rm c}} } \right) - 3 \left(  \dfrac{ {\it x_{\rm 0}} }{  {{\it r}_{\rm c}} } \right)^{-1} + 3 \left( \dfrac{ {{\it x}_{\rm 0}} }{ { {\it r}_{\rm c}} } \right)^{-2}{\rm tan}^{-1} \left(\dfrac{ {{\it x}_{\rm 0}} }{  {{\it r}_{\rm c}} }\right) \right]}{ \left(\dfrac{ {\it x_{\rm 0}} }{ {{\it r}_{\rm c}}} \right)  }, 
%\end{equation}

\begin{eqnarray}
{\it Q}_{\rm sz} &=& \dfrac{\sqrt5\pi\sigma_{\rm T} kT}{4fm_{\rm e}c^2} \times {\it N}_{\rm 0}{\it r}_{\rm c} \times  \nonumber \\
& &  \dfrac{ \left[ {\rm tan}^{-1}\left( \dfrac{ {\it x_{\rm 0}} }{  {{\it r}_{\rm c}} } \right) - 3 \left(  \dfrac{ {\it x_{\rm 0}} }{  {{\it r}_{\rm c}} } \right)^{-1} + 3 \left( \dfrac{ {{\it x}_{\rm 0}} }{ { {\it r}_{\rm c}} } \right)^{-2}{\rm tan}^{-1} \left(\dfrac{ {{\it x}_{\rm 0}} }{  {{\it r}_{\rm c}} }\right) \right]}{ \left(\dfrac{ {\it x_{\rm 0}} }{ {{\it r}_{\rm c}}} \right)  }
\end{eqnarray}
\vspace{10pt}

\parindent=0pt(Pildis, McGaugh 1996). Here, $\sigma_{\rm T}$ is the Thomson scattering cross section and $f$ is a numerical fudge factor owing to the spherical harmonic multipoles; we adopt $f=$8.7 after Pildis and McGaugh (1996). Substituting this formula with  $N_{\rm 0}r_{\rm c}$ mentioned in subsection 4.1, we calculated the upper limit to the CMB quadrupole anisotropy caused by  hot gas in the Local Group. We assume a plasma temperature of 1 keV, a metal abundance of 0.3 solar, $\beta=2/3$, and $x_{\rm 0}=$ 350 kpc. By changing the assumed core radius in the range of 100--500 kpc, the expected CMB anisotropy was calculated, as shown in figure 6. 

\parindent=10pt The results indicate that the expected CMB anisotropy is at most a few nK. This is far below the observed quadrupole anisotropy of 6 $\mu$K. In conclusion, we have not found significant effects caused by hot gas in the Local Group, on either the XRB or CMB\@. 

\bigskip
\parindent=10pt We are grateful to the ASCA team for the operation of ASCA. We thank also the members of the  software team.

\newpage
{\bf References}
\begin{itemize}
\setlength{\itemindent}{-8mm}
\setlength{\itemsep}{-1mm}
\item[] Banday, A. J. \& Gorski,  K. M. 1996, MNRAS, 283, L21
\item[] Bennett, C. L. Kogut, A., Hinshaw, G., Banday, A. J., Wright, E. L., Gorski, K. M., Wilkinson, D. T., Weiss, R.,et al. 1994, ApJ, 436, 423
\item[] Burke, B. E., Mountain, R. W., Daniels, P. J., Dolat, V. S. 1994, IEEE Trans. Nuc. Sci. 41, 375
\item[] Chen, L.-W., Fabian, A. C., \& Gendreau, K. C. 1997, MNRAS, 285, 449
\item[] Davis, D. S., Mulchaey, J. S., \& Mushotzky, R. F. 1999, ApJ, 511, 34
\item[] Fabian, A. C., \& Barcons, X. 1992, ARA\&A, 30, 429
\item[] Fukazawa, Y., Makishima, K., Matsushita, K., Yamasaki, N., Ohashi, T,
Mushotzky, R. F., Sakima, Y., Tsusaka, Y. \& Yamashita, K. 1996, PASJ, 48, 395
\item[] Gendreau, K. C., Mushotzky, R., Fabian, A. C., Holt, S. S., Kii, T., Serlemitsos, P. J., Ogasaka, Y., Tanaka, Y. et al. 1995, PASJ, 47, L5
\item[] Giacconi, R., Rosati, P., Tozzi, P., Nonino, M., Hasinger, G., Norman, C., Bergeron, J., Borgani, S. et al. 2001, ApJ, 551, 624
\item[] Hashinger, G. 1992, in Cambridge University Press: Cambridge, ed. X. Barcons \& A.C. Fabian (  $The Xray Background$), 229
\item[] Hasinger, G., Burg, R., Giacconi, R., Schmidt, M., Tr$\ddot{u}$mper, J. and Zamorani, G. 1998, A\&A, 329,482
\item[] Hickson, P. 1997, ARA\&A, 35, 357
\item[] Inoue, H., Koyama, K., Matsuoka, M., Ohashi,T., Tanaka, Y. 1980, ApJ, 238, 886
\item[] Ishisaki, Y. 1996, PhD Thesis the University of Tokyo
\item[] Ishisaki, Y., Ueda,Y., Kubo, H., Ikebe,Y., Makishima,K., and the GIS team, 1997, ASCA News, No.5
\item[] Kerp, J. 1994, A\&A, 289, 597
\item[] Kerp, J., Burton, W. B., Egger, R., Freyberg, M. J., Hartmann, D., Kalberla, P. M. W., Mebold, U., \& Pietz J. 1999, A\&A, 342, 213
\item[] Kuntz, K. D., \& Snowden, S. L. 2000, ApJ, 543, 195
\item[] Makishima, K., Tashiro, M., Ebisawa, K, Ezawa, H., Fukazawa, Y., Gunji, S., Hirayama, M., Idesawa, E. et al. 1996, PASJ, 48, 171
\item[] Miyaji, T., Ishisaki, Y., Ogasaka, Y., Ueda, Y., Freyberg, M. J., Hasinger, G., \& Tanaka, Y. 1998, A\&A, 334, L13
\item[] Mulchaey, J. S., Davis, D. S., Mushotzy, R. F.,\& Burstein, D. 1996, ApJ, 456, 80
\item[] Mulchaey, J. S., \& Zabludoff A.I. 1998, ApJ, 496, 73
\item[] Mushotzky, R. F., Cowie, L. L., Barger, A. J., Arnaud, K. A. 2000, nature, 404, 459
\item[] Ohashi, T., Ebisawa, K., Fukazawa, Y., Hiyoshi, K., Horii, M., Ikebe, Y., Ikeda, H., Inoue, H. et al. 1996, PASJ, 48, 157
\item[] Parmar, A. N., Guainazzi, M., Oosterbroek, T., Orr, A., Favata, F., Lumb, D., \& Malizia, A. 1999, A\&A, 345, 611
\item[] Peebles, P. J. E. 1990, ApJ, 362, 1
\item[] Pildis, R. A., Bregman, J. N., \& Evrard, A. E. 1995, ApJ, 443, 514
\item[] Pildis, R. A., McGaugh, S. S 1996, ApJ, 470, L77 
\item[] Raymond, J. C., Smith, B. W. 1977, ApJ, 35, 419
\item[] Rocchia, R., Arnaud, A., Blondel, C., Cheron, C., Christy, J. C., Rothenflug, R., Schnopper, H.W., \& Delvaille, J. P. 1984, A\&A, 130, 53
\item[] Serlemitsos, P. J., Jalota, L., Soong, Y., Kunieda, H., Tawara, Y., Tsusaka, Y., Suzuki, H., Sakima, Y., et al. 1995, PASJ, 47, 105
\item[] Sidher, S. D., Sumner, T. J., Quenby, J. J., \& Gambhir, M. 1996, A\&A, 305, 308
\item[] Sidher, S. D., Sumner, T. J., Quenby, J. J. 1999, A\&A, 344, 333
\item[] Snowden, S. L., Freyberg, M. J., Plucinsky, P. P., Schmitt, J. H. M. M. 1995, ApJ, 454, 643
\item[] Snowden, S. L., Egger, R., Freyberg, M. J., McCammon, D., Plucinsky P.P., Sanders, W. T., Schmitt, J. H. M. M., Trumper, J, \& Voges, W. 1997, ApJ, 485, 125
\item[] Suto, Y., Makishima, K., Ishisaki, Y., \& Ogasaka, Y. 1996, ApJ, 461, L33
\item[] Takahashi, I., Fukazawa, Y., Kodaira, K., Makishima, K., Nakazawa, K., \& Xu, H. 2000, PASJ, 52, 769
\item[] Takahashi, T.,Markevitch, M., Fukazawa, Y., Ikebe, Y., Ishisaki, Y., Kikuchi, K., Makishima, K., Tawara, Y., ASCA Image Analysis Working Group  1995, ASCA News, No.3, 34
\item[] Tanaka, Y., Inoue, H., Holt, S. S. 1996, PASJ,  46, L37 .
\item[] Ueda, Y., Takahashi, T., Inoue, H., Tsuru, T., Sakano, M., Ishisaki, Y., Ogasaka, Y., Makishima, K., Yamada, T., \&Ohta, K.  1998, Nature, 391, 866
\item[] Ueda, Y., Takahashi, T., Inoue, H., Tsuru, T., Sakano, M., Ishisaki, Y., Ogasaka, Y., Makishima, K. et al. 1999, ApJ, 518, 656
\end{itemize}

\newpage

\begin{table*}[h]
\begin{center}
\caption{Observation log.}

\begin{tabular*}{18cm}{cccrcc}   \hline\hline
 Field$^\ast$  & Date & Direction & Exposure & Area & $N_{\rm H}$$^{\dagger}$ \\ 
 &   & ($\alpha^{\rm 2000}$, $\delta^{\rm 2000}$) &  (s)  & (deg$^2$) & (10$^{20}$ cm$^{-2}$)  \\ \hline 
LGC & 1996 August 2 to 4 &         & 99002 & 1.60 (1.22)$^{\sharp}$ & 5.1 \\ \hline
(LGC 1) &1996 August 2 & 0$^{\rm h}$52$^{\rm m}$, 36$^{\circ}$30' &  18988 &  0.40 &  5.1 \\
(LGC 2) & 1996 August 3 &  0$^{\rm h}$52$^{\rm m}$, 35$^{\circ}$42' &  27082  &  0.40 & 5.1 \\
(LGC 3) & 1996 August 3 &  0$^{\rm h}$50$^{\rm m}$, 35$^{\circ}$00' &  20136 &  0.40 &  5.1 \\
(LGC 4) & 1996 August 4 & 0$^{\rm h}$48$^{\rm m}$, 34$^{\circ}$18' &  32796  &  0.40 &  5.1\\ \hline
 LSS &  1993 December 26  &  13$^{\rm h}$14$^{\rm m}$, 31$^{\circ}$30'  &   515392  & 7.22 (5.44)$^{\sharp}$ & 1.1 \\ 
  & to 1995 July 8$^{\ddagger}$ &  & & \\  \hline
\end{tabular*}
\end{center}
{\footnotesize
$^{\ast}$ LGC stands for the center of the Local Group; LSS for the reference sky.

$^{\dagger}$ Galactic column density is obtained from EINLINE\@.

$^{\ddagger}$ 76 pointings each for $\sim$20 ks are summed (Ueda et al.\ 1998, 1999).

$^{\sharp}$ Area after removing point sources above a certain threshold flux (see subsection 3.2).
}
\end{table*}

\newpage

\begin{table}[h]
\caption{Result of spectral fits to XRB in the LGC and that in the LSS.}
\begin{center}

\vspace{10pt}
\begin{tabular*}{85mm}{lcc}\hline\hline
 & LGC & LSS \\ \hline
Flux (2--10 keV)$^{\ast}$ in $A E^{-\Gamma_{\rm h}}$ & 4.56$^{+0.67}_{-0.66}$  & 5.58$^{+0.27}_{-0.23}$ \\
        $\Gamma_{\rm h}$ & 1.64$^{+0.16}_{-0.14}$ & 1.46$^{+0.05}_{-0.04}$\\
Flux (0.6--2 keV)$^{\ast}$ in $B E^{-6.0}$ & 0.25$^{+0.41}_{-0.25}$   & 0.52$^{+0.13}_{-0.14}$  \\
$\chi^2/{\rm d.o.f.}^{\dagger}$ & 77.62/70   & 71.28/70     \\ \hline
\end{tabular*}
\end{center}
{\footnotesize
$^\ast$ In unit of $10^{-8}$ erg cm$^{-2}$ s$^{-1}$ sr$^{-1}$.

$^\dagger$ Systematic errors are not included.}
\end{table}

\newpage

\begin{figure}[h]
 \begin{center}
%\psbox[height=8.5cm,width=8.5cm]{./eps/coordinate2.eps}
\psbox[height=8.5cm,width=8.5cm]{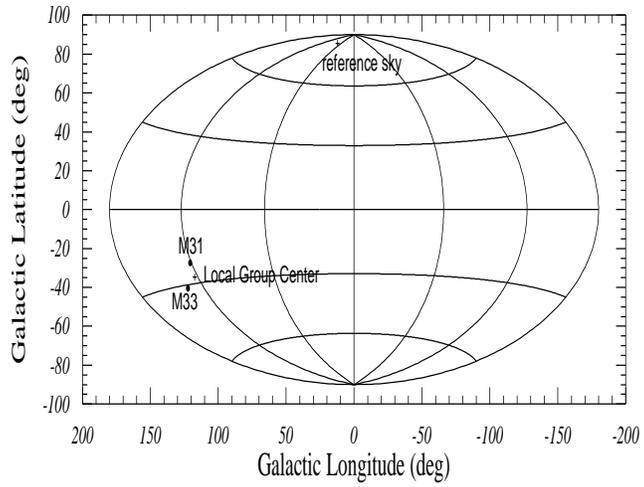}
 \hspace{20pt}
 \end{center}
 \caption{Positions of the observed fields (LGC) and the reference fields (LSS) in the Galactic coordinates, both marked with plus. The positions of M 31 and M 33 are also shown.}
\end{figure}

\newpage

\begin{figure}[h]
 \begin{center}
 \psbox[height=8.5cm,width=8.5cm]{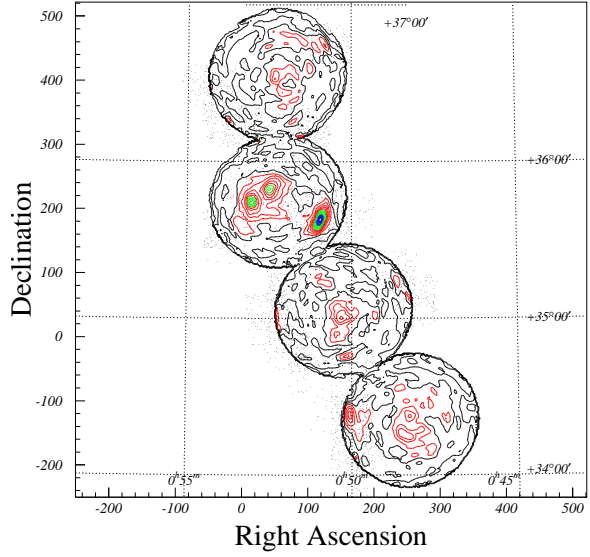}
 \end{center}
 \caption{NXB-subtracted and smoothed image of the LGC in the 0.7$-$7.0 keV band. The brightness is plotted in a linear scale. The dotted areas show the point source masks, of which the threshold level is 2$\times$10$^{-13}$ erg s$^{-1}$ cm$^{-2}$ in the 2$-$10 keV band.}
\end{figure}

\newpage

\begin{figure*}[h]
 \begin{center}
\psbox[height=8.5cm,width=8.5cm]{980929-1-3.eps01}
\psbox[height=8.5cm,width=8.5cm]{980929-4-3.eps01}
 \end{center}
 \caption{The XRB spectra in the LGC (left) and LSS (right) fields, fitted with two power-laws. The photon index of the soft power-law is fixed at 6.0, and the Galactic absorption is applied to the hard power-law. Systematic errors are not included.}
\end{figure*}

\newpage

\begin{figure}[h]
 \begin{center}
 \psbox[height=8.5cm,width=8.5cm]{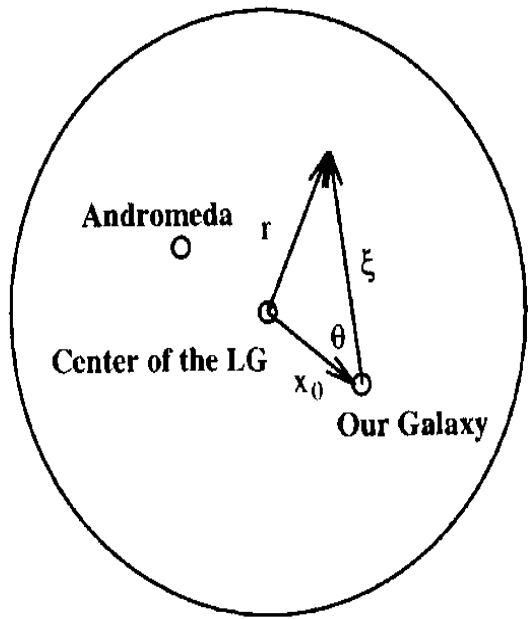}
 \end{center}
\caption{Geometry of the Local group halo, quoted from Suto et al.\ (1996).}
\end{figure}

\newpage

\begin{figure}[h]
 \begin{center}
\psbox[height=8.5cm,width=8.5cm]{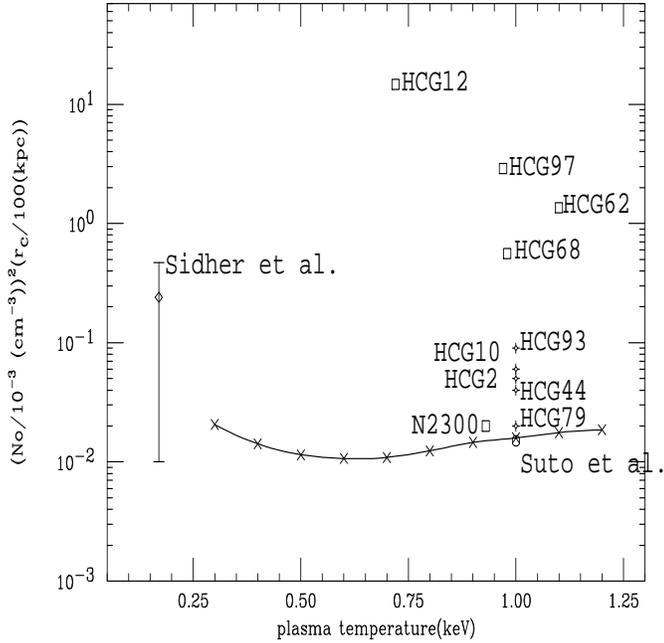}
 \end{center}
 \caption{Upper limits of the quantity ${N_{\rm 0}}^2 r_{\rm c}$ of the Local Group, plotted against a plasma temperature, assuming $\beta=2/3$, a metal abundance 0.3 solar, and the distance to center of Local Group 350 kpc (crosses). A diamond represents the ROSAT result (Sidher et al. 1999), assuming a core radius of 150 kpc and a gas temperature of 0.17 keV\@. The circle indicates the expected value for the Local Group halo calculated by Suto et al.\ (1996). The squares represent the parameters for some of the X-ray detected galaxy groups, and stars indicate the upper limits for the X-ray undetected groups, assuming a gas temperature of 1.0 keV, extracted from Pilidis and McGaugh (1996).}
\end{figure}

\newpage

\begin{figure}[h]
\begin{center}
 \psbox[height=8.5cm,width=8.5cm]{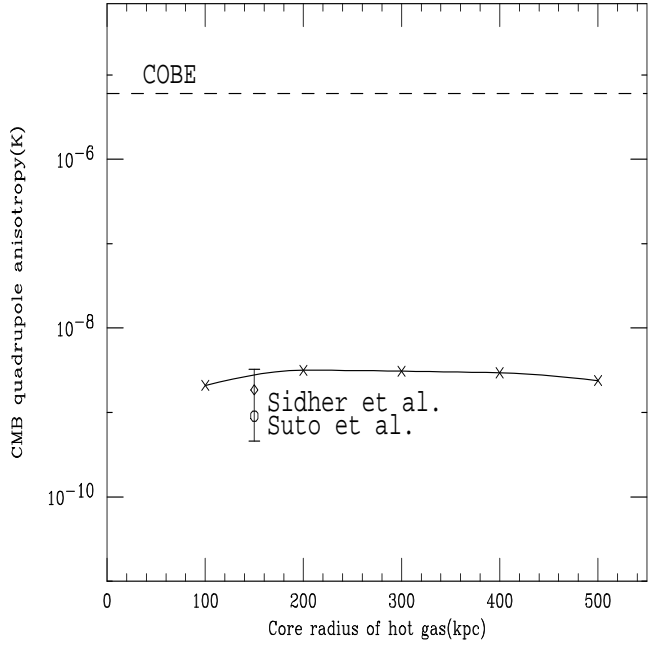}
\end{center}
\caption{Upper limits of the CMB quadrupole anisotropy caused by hot gas in the Local Group, calculated as a function of the assumed halo core radius, assuming a gas temperature of 1.0 keV, $\beta=2/3$, a metal abundance of 0.3 solar and the distance to center of the Local Group as being 350 kpc (crosses). The circle indicates calculation by Suto at al.\ (1996). The diamond indicates the ROSAT result (Sidher et al.\ 1999), assuming a core radius of 150 kpc and a gas temperature of 0.17 keV\@. The dotted line is the CMB quadrupole anisotropy observed by COBE (Bennett et al.\ 1994). }
\end{figure}

\end{document}